\documentclass[fp,twocolumn]{jpsj3}

\usepackage{txfonts}
\usepackage[dvipdfmx]{graphicx}
\usepackage{mathrsfs} 
\usepackage[mathscr]{euscript}
\usepackage{amsmath,amssymb}
\usepackage{bm}
\usepackage{color}

\def\ket#1{ | #1 \rangle }
\def\bra#1{ \langle #1 | }

\def\mod{{\rm mod}}

\title{Infinite-Size Density Matrix Renormalization Group with Parallel Hida's Algorithm}

\author{Hiroshi Ueda \thanks{h\_ueda@riken.jp}}
\inst{Computational Materials Science Research Team, RIKEN Center for Computational Science (R-CCS), Kobe, Hyogo 650-0047, Japan} 

\abst{In this study, we report a parallel algorithm for the infinite-size density matrix renormalization group (iDMRG) that is applicable to one-dimensional (1D) quantum systems with $\ell$-site periods, where $\ell$ is an even number. It combines Hida's iDMRG applied to random 1D spin systems with a variant of McCulloch's wavefunction prediction. This allows us to apply $\ell/2$ times the computational power to accelerate the investigation of multileg frustrated quantum systems in the thermodynamic limit, which is a challenging simulation. We performed benchmark calculations for a spin-1/2 Heisenberg model on a YC8 kagome cylinder using the parallel iDMRG. It was found that the proposed iDMRG was efficiently parallelized for shared memory and distributed memory systems, and provided bulk physical quantities such as total energy, bond strength on nearest-neighbor spins, and spin--spin correlation functions and their correlation lengths without finite-size effects. Moreover, the variant of the wavefunction prediction increased the speed of Lanczos methods in the parallel iDMRG by approximately three times.}

\begin{document}
\maketitle

\section{Introduction}
Quantum spin systems on low-dimensional lattices with geometrical frustration, which are beyond the reach of quantum Monte Carlo simulations, are a fascinating subject of study in condensed matter physics because a variety of nontrivial quantum phases can emerge due to the coexistence of geometrical frustration and quantum fluctuation. The density matrix renormalization group (DMRG) proposed by White is a powerful tool for analyzing low-energy states of such systems~\cite{White:PRL69-PRB48}. In particular, the infinite-size DMRG (iDMRG) can be used to directly investigate one-dimensional (1D) systems in the thermodynamic limit~\cite{White:PRL69-PRB48} and has been extensively studied in condensed matter physics~\cite{RevModPhys.77.259, 201196}. The iDMRG has recently been applied as a useful detector for finding symmetry-protected topological phases in 1D systems, and this has enhanced its importance~\cite{PhysRevB.81.064439,PhysRevB.83.035107,PhysRevB.84.235128,PhysRevB.86.125441}. Therefore the sophistication of numerical algorithms for iDMRG has also become important. 

One of the goals of iDMRG is to obtain a wavefunction represented by a translationally invariant matrix product state~\cite{affleck1988, fannes1992, Ostlund:PRL75, Rommer:PRB55} (MPS) with a unit cell. To achieve this, two algorithm are typically used to accelerate the iDMRG, the product wavefunction renormalization group (PWFRG)~\cite{JPSJ.64.4084,JPSJ.75.014003,JPSJ.77.114002} and McCulloch's wavefunction prediction~\cite{0804.2509,JPSJ.79.044001}, when the translationally invariant MPS has a unit cell consisting of few sites. 

In contrast, a generalization of the iDMRG for position-dependent Hamiltonians was proposed by Hida~\cite{Hida:JPSJ65} and applied to the analysis of 1D quantum random systems~\cite{doi:10.1143/JPSJ.66.3237,doi:10.1143/JPSJ.66.330,PhysRevLett.83.3297}. It has been claimed~\cite{Hida:JPSJ65} that Hida's iDMRG is useful for studying systems with large unit cells and can be accelerated by wavefunction prediction methods~\cite{JPSJ.64.4084}. However, we still cannot implement it on multileg ladder/cylinder systems, typically more than 10 legs, which have recently become the typical target of the DMRG~\cite{doi:10.1146/annurev-conmatphys-020911-125018}. 

In this article, we propose an extension of the iDMRG with a variant of McCulloch's wavefunction prediction that can be applied to quantum systems with large ($\ell$-site) unit cells. We show that our algorithm is efficiently parallelized for both shared memory and distributed memory systems and that the wavefunction prediction reduces the number of Lanczos iterations to approximately a third of that without the prediction. Moreover, this method is compatible with the subtraction method~\cite{doi:10.1146/annurev-conmatphys-020911-125018} and can easily obtain the total energy in the bulk limit under a fixed $m$, which is the number of states maintained for block-spin variables. The numerical accuracy of the iDMRG can be considered by using the truncation error or discarded weight $\varepsilon$ as a function of $m$~\cite{White:PRL69-PRB48}, where the results of the iDMRG for multileg systems are strongly dependent on $\varepsilon$. The error $\varepsilon$ is not given uniquely in the MPS with multisite unit cells, and we succeeded in finding an appropriate $\varepsilon$ for our parallel iDMRG that significantly suppresses higher-order terms of $\varepsilon$ in the bulk energy with respect to $\varepsilon$. We can apply $\ell/2$ times the computational power to challenging simulations, thus accelerating the examination of multileg frustrated quantum systems in the thermodynamic limit through our parallel algorithm. We applied them to the spin-1/2 Heisenberg model on a YC8 kagome cylinder~\cite{Yan03062011}, which can be mapped to 1D quantum systems with a $12$-site unit cell. 

The remainder of this article is organized as follows. In the next section, we recall the algorithm for Hida's iDMRG in terms of the formalism of the matrix product. In Sec. \ref{sec.p_idmrg}, we introduce our proposed algorithm for the parallel iDMRG. We test the performance of the iDMRG in Sec. \ref{sec.bc}, where we show the effectiveness of a variant of wavefunction prediction~\cite{0804.2509} to reduce the number of Lanczos iterations in the parallel iDMRG, and introduce an appropriate value of $\varepsilon$ for extrapolations to estimate the bulk energy of the YC8 kagome cylinder. We also discuss the bond strength of nearest-neighbor spins, and a spin-spin correlation function and its correlation length. We summarize our conclusions in the final section, where we state the relation between Hida's iDMRG and the real-space parallel DMRG~\cite{PhysRevB.87.155137}. 

\section{Hida's iDMRG from the Perspective of Matrix Product Formalism}
\label{sec.hida}

\begin{figure}[htb]
\centering
\includegraphics[width=8.5cm]{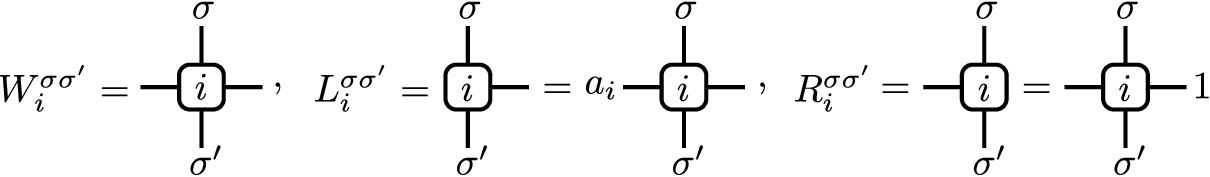} 
\caption{Graphical representations of $W^{\sigma \sigma'}_{i}$, $L^{\sigma \sigma'}_{i}$, and $R^{\sigma \sigma'}_{i}$, where vertical and horizontal lines emerging from rounded squares represent physical and auxiliary variables, respectively. Integer $a_i$ represents the number of rows (columns) of $W^{\sigma \sigma'}_{i}$ ($W^{\sigma \sigma'}_{i-1}$). The number of lines without indices of states represents the rank of the tensor, which means that $W^{\sigma \sigma'}_{i}$ is a matrix and $L^{\sigma \sigma'}_{i}$ and $R^{\sigma \sigma'}_{i}$ are row and column vectors, respectively.}
\label{fig:diagram1}
\end{figure}

In this section, we review the algorithm of Hida's iDMRG~\cite{Hida:JPSJ65} in terms of a matrix product formalism. This algorithm targets $\ell$-site systems represented by position-dependent Hamiltonians, where $\ell$ is an even number. Using the formalism of the matrix product operator~(MPO)~\cite{0804.2509}, we can express a position-dependent Hamiltonian as 
\begin{equation}
h_{\bm{\sigma} \bm{\sigma}'} = L^{\sigma^{~}_1 \sigma'_1}_{1} 
\left[ \prod_{i=2}^{\ell-1} W^{\sigma^{~}_i \sigma'_i}_{i} \right]
R^{\sigma^{~}_{\ell} \sigma'_{\ell}}_{\ell}~, \label{eq.mpo}
\end{equation}
where $\bm{\sigma}=(\sigma_1,\cdots,\sigma_\ell)$ and $W^{\sigma^{~}_i \sigma'_i}_{i}$ is a lower-triangular matrix defined for the outer product of local states $\ket{\sigma^{~}_i}\bra{\sigma'_i}$ for the $i$th site. The left and right boundary vectors, $L^{\sigma^{~}_i \sigma'_i}_{i}$ and $R^{\sigma^{~}_i \sigma'_i}_{i}$, are identical to the last row and first column of the matrix $W^{\sigma^{~}_i \sigma'_i}_{i}$, respectively. 
Hereinafter, unless otherwise noted, we abbreviate the subscripts in $\sigma^{~}_i$ for the sake of simplicity. Figure \ref{fig:diagram1} shows a graphical representation of $W^{\sigma \sigma'}_{i}$, $L^{\sigma \sigma'}_{i}$, and $R^{\sigma \sigma'}_{i}$. 
\begin{figure}[htb]
\centering
\includegraphics[width=8.5cm]{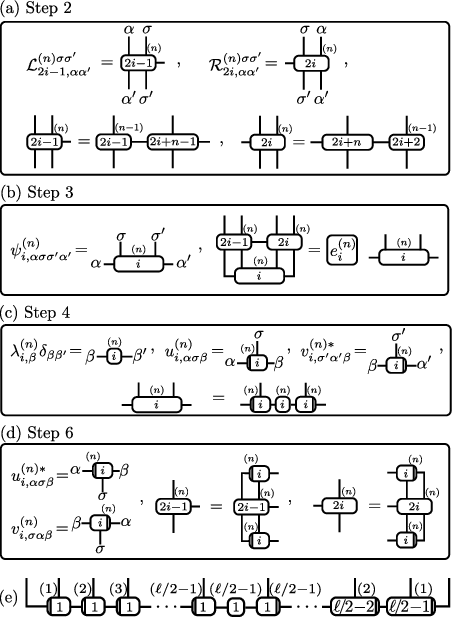} 
\caption{Graphical representations of the tensor contractions in (a) Step 2, (b) Step 3, (c) Step 4, (d) Step 6, and (e) Eq.~(\ref{eq.hida_mps}) in Hida's iDMRG algorithm, where $\delta_{\beta\beta'}$ is Kronecker's delta. We calculate the sum of configurations of links connecting neighboring diagrams.}
\label{fig:diagram2}
\end{figure}

Hida's iDMRG proceeds as follows:
\begin{enumerate}
\item Give $W^{\sigma \sigma'}_{i}$ in Eq.~(\ref{eq.mpo}) and prepare pairs of blocks, $L^{(0)}_{2i-1} = L^{~}_{2i-1}$ and $R^{(0)}_{2i} = R^{~}_{2i}$, where $1 \leq i \leq \ell/2$. Then, set the number of iterations to $n=1$. 

\item Expand each pair of blocks as follows:
\begin{eqnarray}
\mathcal{L}^{(n)\sigma \sigma'}_{2i-1,\alpha \alpha'} & = & L^{(n-1)\alpha\alpha'}_{2i-1} W^{\sigma\sigma'}_{2i+n-1}~, \label{eq.ebl} \\
\mathcal{R}^{(n)\sigma \sigma'}_{2i,\alpha \alpha'} & = & W^{\sigma\sigma'}_{2i+n} R^{(n-1)\alpha\alpha'}_{2i+2}~, \label{eq.ebr}
\end{eqnarray}
where $1 \leq i \leq \ell/2-n$ and $1 \leq \alpha \leq \min[d^n,m]$, where $d$ is the number of degrees of freedom of the local state~[Fig.~\ref{fig:diagram2}(a)]. 

\item Solve the eigenvalue problem for each superblock Hamiltonian~\cite{White:PRL69-PRB48} $H^{(n)}_i = \{ \mathcal{L}^{(n)\sigma \sigma''}_{2i-1,\alpha \alpha''} \mathcal{R}^{(n)\sigma' \sigma'''}_{2i, \alpha' \alpha'''} \}$ as $H^{(n)}_{i} \Psi^{(n)}_{i} = e^{(n)}_{i} \Psi^{(n)}_{i}$ by an iterative method---for example, Lanczos, Jacobi--Davidson, etc., where the ground-state energy and a corresponding eigenvector are represented by $e^{(n)}_{i}$ and $\Psi^{(n)}_{i}= \{ \psi^{(n)}_{i,\alpha \sigma \sigma' \alpha'} \}$, respectively~[Fig.~\ref{fig:diagram2}(b)]. An initial vector $\tilde{\Psi}^{(n)}_{i}$ is required to start the iteration method, which is often given randomly.

\item Apply singular value decomposition (SVD) to $ \Psi^{(n)}_{i}$ as
$\psi^{(n)}_{i,\alpha \sigma,\sigma' \alpha'} = \left( U^{(n)}_{i} \Lambda^{(n)}_{i} V^{(n)\dagger}_{i} \right)_{\alpha \sigma, \sigma' \alpha'}$,
where $U^{(n)}_{i}=\{u^{(n)}_{i,\alpha \sigma, \gamma}\}$ and $V^{(n)}_{i}=\{v^{(n)}_{i,\sigma'\alpha',\gamma}\}$ with $1 \leq \gamma \leq \min[d^{n+1},md]$ are unitary matrices. The diagonal matrix $\Lambda^{(n)}_i={\rm diag} ( \{\lambda^{(n)}_{i,\gamma}\} )$ contains singular values and is normalized as $\sum_\gamma (\lambda^{(n)}_{i,\gamma} )^2 = 1$, where $\lambda^{(n)}_{i,1} \leq \lambda^{(n)}_{i,2}\leq \cdots$, because $\Psi^{(n)\dagger}_{i}\Psi^{(n)}_{i} = 1$~[Fig.~\ref{fig:diagram2}(c)]. 

\item If $n = \ell/2 - 1$, complete the calculations.

\item Apply block-spin transformations, as depicted in Fig.~\ref{fig:diagram2}(d), to each expanded block as follows:
\begin{eqnarray}
L^{(n)\beta\beta'}_{2i-1} & = & \sum_{\alpha \sigma \alpha' \sigma'} u^{(n)*}_{i,\alpha \sigma \beta} \mathcal{L}^{(n)\sigma\sigma'}_{2i-1, \alpha \alpha'} u^{(n)}_{i,\alpha' \sigma' \beta'}~, \label{eq.L} \\
R^{(n)\beta\beta'}_{2i} & = & \sum_{\sigma \alpha \sigma' \alpha'} v^{(n)}_{i,\sigma \alpha \beta } \mathcal{R}^{(n)\sigma\alpha'}_{2i,\alpha\sigma'} v^{(n)*}_{i,\sigma' \alpha' \beta'}~, \label{eq.R}
\end{eqnarray}
where $1 \leq \beta \leq \min[d^{n+1},m]$. The truncation of the number of degrees of freedom of the blocks can be introduced in this step. 

\item Set $n+1 \rightarrow n$ and go to Step 2. 

\end{enumerate}
Through these processes, we obtain a variational/exact ground state $\Psi=\{ \psi_{ \bm{\sigma} } \}$ of the original Hamiltonian as follows: 
\begin{eqnarray}
\psi_{ \bm{\sigma} } & = & \mathcal{U}^{(1)}_{1,\sigma^{~}_1\sigma^{~}_2} \mathcal{U}^{(2)}_{1,\sigma^{~}_{3}} \mathcal{U}^{(3)}_{1,\sigma^{~}_{4}} \cdots \mathcal{U}^{(\ell/2-1)}_{1,\sigma^{~}_{\ell/2}} \Lambda^{(\ell/2-1)}_{1} \nonumber \\
&& \times \mathcal{V}^{(\ell/2-1)\dagger}_{1,\sigma^{~}_{\ell/2+1}} \cdots \mathcal{V}^{(2)\dagger}_{\ell/2-2,\sigma^{~}_{\ell-2}}
\mathcal{V}^{(1)\dagger}_{\ell/2-1,\sigma^{~}_{\ell-1} \sigma^{~}_\ell} ~,
\label{eq.hida_mps}
\end{eqnarray}
where $\mathcal{U}^{(i-1)}_{1,\sigma^{~}_{i}} = \{ u^{(i-1)}_{1,\alpha, \sigma_{i}, \beta} \}$ and $\mathcal{V}^{(\ell-i)}_{i-\ell/2,\sigma^{~}_{i}} = \{ v^{(\ell-i)}_{i-\ell/2, \sigma_{i}, \alpha, \beta} \}$ are matrices defined for the local state $\ket{\sigma_i}$. In Eq.~(\ref{eq.hida_mps}), $\mathcal{U}^{(1)}_{1,\sigma^{~}_{1}\sigma^{~}_{2}} = \{ u^{(1)}_{1,\sigma_1 \sigma_{2}, \beta} \}$ and $\mathcal{V}^{(1)}_{\ell/2-1,\sigma^{~}_{\ell-1}\sigma^{~}_{\ell}} = \{ v^{(1)}_{\ell/2-1, \sigma^{~}_{\ell-1} \sigma^{~}_{\ell}, \beta} \}$ are column and row vectors defined for $\ket{\sigma_1\sigma_2}$ and $\ket{\sigma^{~}_{\ell-1} \sigma^{~}_{\ell}}$, respectively [Fig.~\ref{fig:diagram2}(e)]. 

\begin{figure}[htb]
\centering
\includegraphics[width=8.5cm]{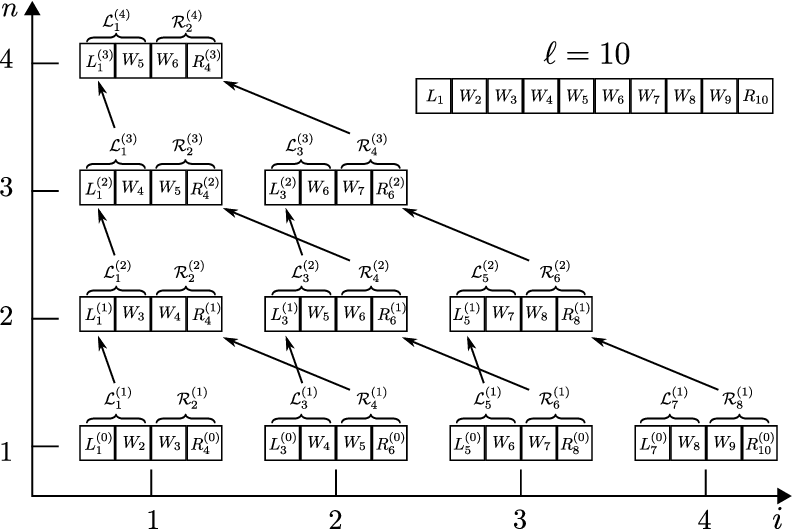} 
\caption{ Schematic procedure of Hida's iDMRG for 10-site systems. The overbraces and arrows represent expanding pairs of blocks in Step 2 and block-spin transformations in Step 6, respectively.  }
\label{fig:Hida_idmrg}
\end{figure}

For the overall picture of Hida's iDMRG, we show a schematic procedure for $\ell=10$ in Fig. \ref{fig:Hida_idmrg}. 
Critical to this algorithm are preparing and growing $\ell/2-1$ pairs of blocks $L^{(n)}_{2i-1}$ and $R^{(n)}_{2i}$ to provide a suitable environment in each position-dependent DMRG calculation. Because of this careful treatment, this infinite-size algorithm can be effectively applied to analyze the ground states of random quantum 1D systems.

Moreover, the process of expanding and block-spin transformations for each pair of $L^{(n)}_{2i-1}$ and $R^{(n)}_{2i}$ can be parallelized easily. One-to-one communications between nearest-neighbor nodes are required for $R^{(n-1)}_{2(i+1)}$ in Eq.~(\ref{eq.ebr}), and the cost per node is constant irrespective of $\ell$. This property is suitable for message-passing interface-parallel (MPI) programming. 

\section{Parallel ${\rm {\bf i}}$DMRG Algorithm for Systems with $\ell$-Site Periodic Structure} \label{sec.p_idmrg}

In this section, we describe a combination of Hida's iDMRG~\cite{Hida:JPSJ65} with a variant of McCulloch's wavefunction prediction~\cite{0804.2509,JPSJ.79.044001}. A target Hamiltonian containing $k\ell~(k \gg 1)$ sites can be represented by an MPO as
\begin{equation}
h_{\bm{\sigma} \bm{\sigma}'} = \left[ \prod_{j=0}^{k-1} \prod_{i=1}^{\ell} W^{\sigma^{~}_{j\ell+i} \sigma'_{j\ell+i} }_{i} \right]_{a_1,1}~. \label{eq.mpo2}
\end{equation}

We can construct a parallel iDMRG with wavefunction predictions by replacing Steps 2, 3, and 5 of Hida's iDMRG in the previous section with the following procedures, 
\begin{itemize}
\item[2.] Expand each pair of blocks as
\begin{eqnarray}
\mathcal{L}^{(n)\sigma \sigma'}_{2i-1,\alpha \alpha'} & = & L^{(n-1)\alpha\alpha'}_{2i-1} W^{\sigma\sigma'}_{f(2i+n-1)} ~,\\
\mathcal{R}^{(n)\sigma \sigma'}_{2i,\alpha \alpha'} & = & W^{\sigma\sigma'}_{f(2i+n)} R^{(n-1)\alpha\alpha'}_{2(i+1)} ~,
\end{eqnarray}
where $f(k) = 1 + {\rm mod}[k-1,\ell]$ and $R^{(n)}_{\ell + 2} = R^{(n)}_{2}$. The range of $i$ is always $1 \leq i \leq \ell/2$.
\item[3.] The initial vector $ \tilde{\Psi}^{(n)}_{i} = \{ \tilde{\psi}^{(n)}_{i\alpha \sigma \sigma' \alpha'} \}$ for iteration methods is a random vector if $n=1$ and $2$. When $n \geq 3$, as shown in Fig.~\ref{fig:diagram3}, $ \tilde{\Psi}^{(n)}_{i} $ is given by wavefunction prediction methods~\cite{0804.2509, JPSJ.79.044001} as follows:
\begin{eqnarray}
\tilde{u}^{(n-1)}_{i\beta \sigma' \alpha'} & = & \left\{ \begin{matrix}
0 & (\lambda^{(n-2)}_{i\beta} = 0) \\
\left( \lambda^{(n-2)}_{i\beta} \right)^{-1} u^{(n-1)}_{i\beta \sigma' \alpha'} \lambda^{(n-1)}_{i\alpha'} & (\lambda^{(n-2)}_{i\beta} > 0)
\end{matrix} \right. ~, \nonumber \\
\tilde{\psi}^{(n)}_{i\alpha \sigma \sigma' \alpha'} & = & \sum_{\beta} \lambda^{(n-1)}_{i\alpha} 
v^{(n-1)*}_{i\alpha \sigma \beta} \tilde{u}^{(n-1)}_{i+1,\beta \sigma' \alpha'}~. \label{eq:wfprediction}
\end{eqnarray}
Using $\tilde{\Psi}^{(n)}_{i}$, solve an eigenvalue problem of the Hamiltonian of each superblock.
\begin{figure}[htb]
\centering
\includegraphics[width=8.5cm]{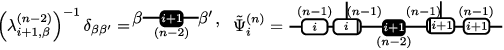} 
\caption{Graphical representations of tensor contractions in Eq.~(\ref{eq:wfprediction}).}
\label{fig:diagram3}
\end{figure}
\item[5.] If $\mod [n+1,\ell]=0$, estimate the ground-state energy per site as $e^{(n)}_{\rm g} = \left( e^{(n)}_1 - e^{(n-\ell)}_{1} \right)/2\ell$ to subtract boundary effects~\cite{doi:10.1146/annurev-conmatphys-020911-125018}, where $e^{(-1)}_{1} = 0$. 
Then, if $e^{(n)}_{\rm g}$ converges with respect to $n$, complete the iDMRG calculation.
\end{itemize}

\begin{figure}[htb]
\centering
\includegraphics[width=8.5cm]{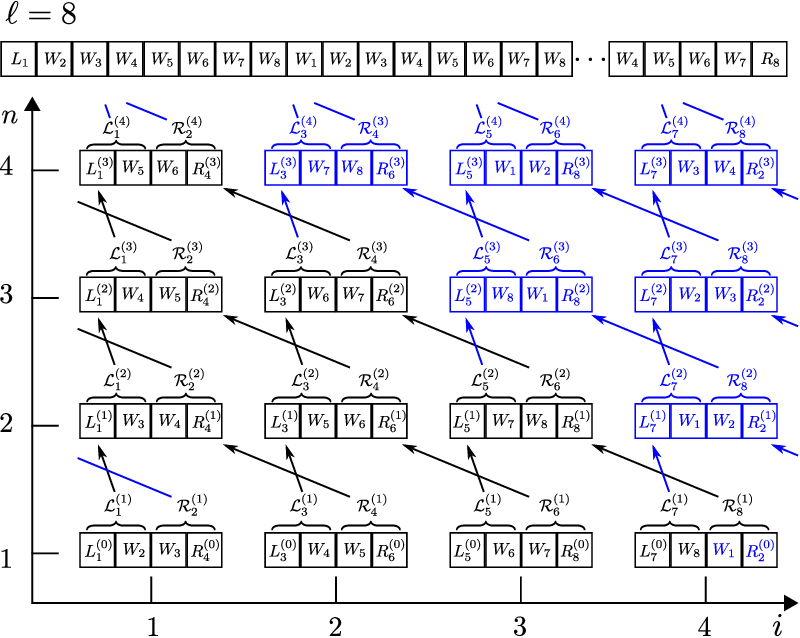} 
\caption{ (Color online) Schematic procedure of the parallel iDMRG for a system with an eight-site period. The blue parts highlight the difference between the parallel iDMRG for eight-site period structures and Hida's iDMRG 10-site period structures in Fig. \ref{fig:Hida_idmrg}. }
\label{fig:parallel_idmrg}
\end{figure}

As shown in Fig.~\ref{fig:parallel_idmrg}, this parallel iDMRG for systems with $\ell$-site periods can be implemented by introducing slight modifications to the Hida's iDMRG for $\ell+2$-site stems.

\begin{figure}[htb]
\centering
\includegraphics[width=8.5cm]{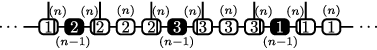} 
\caption{Graphical representation of a unit of the uniform part of the matrix product structures in Eq.~(\ref{eq.mps2}) for $\ell = 6$.}
\label{fig:diagram4}
\end{figure}

Following the calculations, we obtain an MPS for the ground state of the original Hamiltonian using the wavefunction prediction method iteratively; namely,
\begin{eqnarray}
\psi_{ \{ \sigma_i \} } & = & \mathcal{U}^{(1)}_{1,\sigma^{~}_1\sigma^{~}_2} \mathcal{U}^{(2)}_{1,\sigma^{~}_{3}} \mathcal{U}^{(3)}_{1,\sigma^{~}_{4}} \cdots \mathcal{U}^{(n)}_{1,\sigma^{~}_{n+1}} \Lambda^{(n)}_{1} \nonumber \\
&& \times \left[ \prod_{j=0}^{k-1-\frac{n+1}{\ell/2}} \prod_{i=1}^{\ell/2} \mathcal{V}^{(n)\dagger}_{i,\sigma_{n+j\ell+2i}} \tilde{\mathcal{U}}^{(n)}_{i+1,\sigma_{n+j\ell+2i+1}} \right] \nonumber \\
&& \times \mathcal{V}^{(n)\dagger}_{1,\sigma^{~}_{k\ell-n}} \cdots \mathcal{V}^{(2)\dagger}_{\ell/2-2,\sigma^{~}_{k\ell-2}}
\mathcal{V}^{(1)\dagger}_{\ell/2-1,\sigma^{~}_{k\ell-1} \sigma^{~}_{k\ell}} ~, \label{eq.mps2}
\end{eqnarray}
where $\tilde{\mathcal{U}}^{(n)}_{1,\sigma} = \{ \tilde{u}^{(n)}_{1,\alpha, \sigma, \beta} \}$ and $\tilde{\mathcal{U}}^{(n)}_{\ell/2+1} = \tilde{\mathcal{U}}^{(n)}_{1}$ (Fig.~\ref{fig:diagram4}). %

\section{Benchmark Calculations}
\label{sec.bc}

To test the numerical performance of our parallel iDMRG, we estimate the total energy, bond strength on nearest-neighbor spins, spin-spin correlation functions, and correlation lengths of the spin-1/2 Heisenberg model on a YC8 kagome cylinder of infinite length. The Hamiltonian is given as
$\mathcal{H} = \sum_{\langle i,j \rangle} \bm{s}_i \cdot \bm{s}_j$,
where the sum runs over nearest-neighbor sites. The shape of the cylinder YC8 is shown in the inset of Fig.~\ref{fig:t_vs_x}. The Hamiltonian of the cylinder can be represented by an MPO with $\ell=12$ as in Eq.~(\ref{eq.mpo2}).  
The ground state of this model has been widely studied using the finite-size DMRG~\cite{Yan03062011, PhysRevLett.109.067201,PhysRevB.91.104418}. We show that the parallel iDMRG can estimate consistent physical quantities using $m$ only up to $2800$. In this paper, we do not introduce block diagonalizations with respect to typical quantum numbers, for example, the total spin and its $z$ component. Of course, our parallel iDMRG is compatible with the use of abelian and non-abelian symmetries~\cite{0295-5075-57-6-852,PhysRevB.97.134420,Weichselbaum2018}.

\subsection{Parallel performance}
\begin{figure}[htb]
\centering
\includegraphics[width=8.5cm]{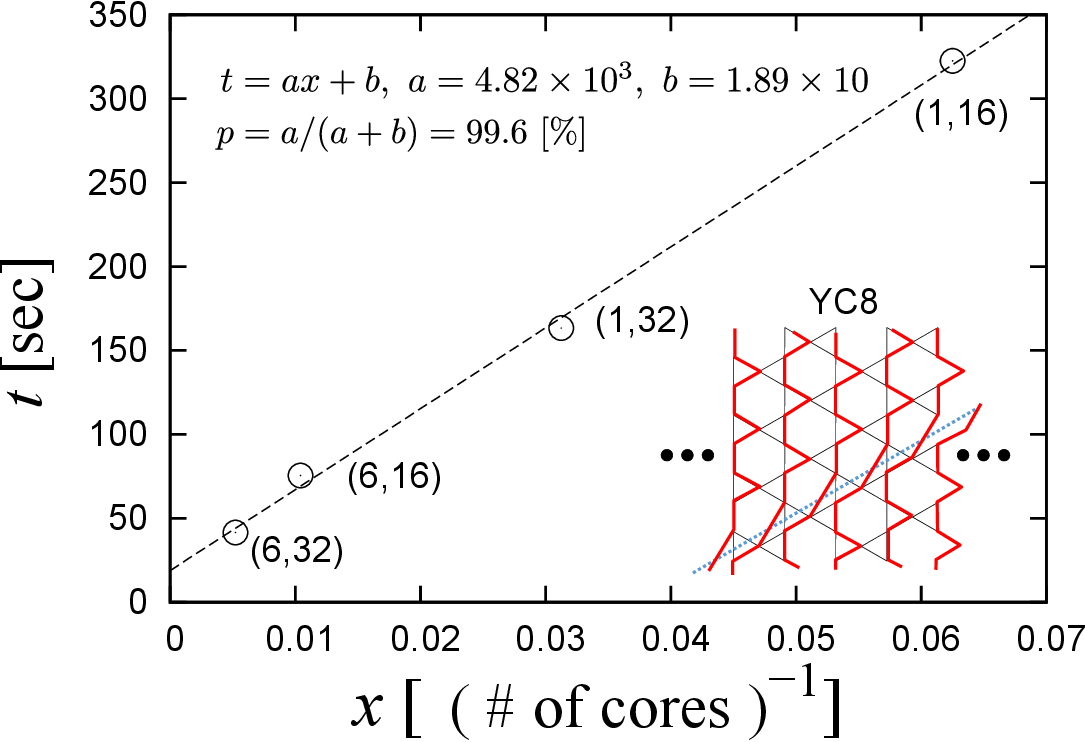} 
\caption{(Color online) Parallel performance of our iDMRG while maintaining $m=1000$ states. Computations were performed by using 32-core Fujitsu SPARC64 Xlfx 1.975 GHz nodes. Each pair of numbers $(q,r)$ located nearby each plot respectively indicates the numbers of nodes and threads per node in the hybrid MPI/OpenMP parallel calculations. Calculation time $t$ is defined by the average time of parallel iDMRGs per iteration, where the number of iterations $n$ is up to $1200$. In the inset, the paths of the matrix product of our MPO and MPS are denoted by bold red lines on the YC8 cylinder, where the blue broken line separates the MPO into periods of the MPO.}
\label{fig:t_vs_x}
\end{figure}

We first evaluated the parallel performance of our iDMRG as shown in Fig.~\ref{fig:t_vs_x}. The time for the calculation $t$ was fitted by a linear function $t = ax + b$, where $x=(qr)^{-1}$, $a$ and $b$ are the reciprocal of the parallel cores, and the parallel and serially processed parts of our calculations, respectively. Parallel efficiency $p$ is defined by $a/(a+b)$, and we obtained $p \sim 99.6~[\%]$ in the calculations. This means that the parallelization worked well up to several hundred cores in this system. 

\subsection{Effect of wavefunction prediction}
\begin{figure}[htb]
\centering
\includegraphics[width=8.5cm]{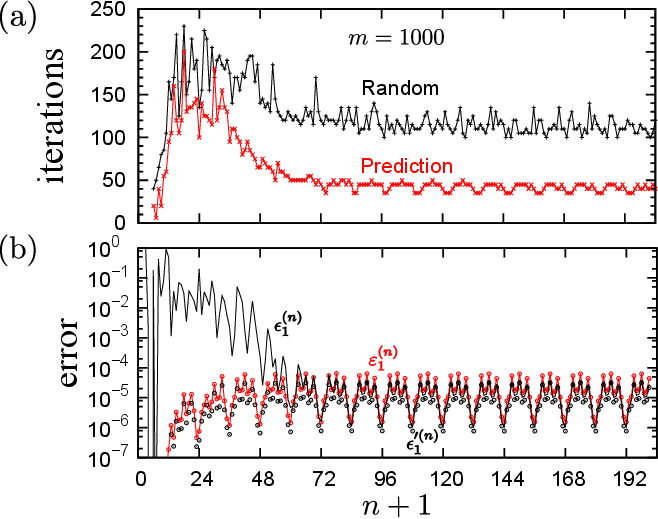} 
\caption{(Color online) (a) Number of iterations for the Lanczos method with/without wavefunction prediction (+/\textcolor{red}{$\times$}) versus the number of iterations for the parallel iDMRG. (b) The fidelity errors $\epsilon^{(n)}_{1}$ in Eq.~(\ref{eq.fid1}) and $\epsilon^{\prime(n)}_{1}$ in Eq.~(\ref{eq.fid2}) and the truncation error $\varepsilon^{(n)}_{1}$ in Eq.~(\ref{eq.tru}) are denoted by black lines, black circles, and red lines with circles, respectively. }
\label{err_vs_itr.eps}
\end{figure}

The wavefunction prediction methods in Step 3 of the parallel iDMRG are used to accelerate iteration methods for eigenvalue problems. The degree of acceleration when solving problems using prediction can be discussed using the fidelity error
\begin{equation}
\epsilon^{(n)}_{i} = 1 - \frac{ | \tilde{\Psi}^{(n)\dagger}_{i} \Psi^{(n)}_{i} | }{ || \tilde{\Psi}^{(n)}_{i} || }~. \label{eq.fid1}
\end{equation}
As the Schmidt rank of $\tilde{\Psi}^{(n)}_{i}$ is up to $m$, the fidelity error $\epsilon^{(n)}_{i}$ must not be less than the best fidelity error $\epsilon^{\prime(n)}_{i}$, given by
\begin{equation}
\epsilon^{\prime(n)}_{i} = 1 - \frac{ \Phi^{(n)\dagger}_{i} \Psi^{(n)}_{i} }{ || \Phi^{(n)}_{i} || } = 1- \sqrt{ \sum_{\beta=1}^{m} \left( \lambda^{(n)}_{i,\beta} \right)^2 }, \label{eq.fid2}
\end{equation}
where $\Phi^{(n)}_i$ is an approximated eigenvector defined by $\Phi^{(n)}_i = \{ \sum_{\beta=1}^{m} u^{(n)}_{i,\alpha \sigma \beta} \lambda^{(n)}_{i,\beta} v^{(n)*}_{i,\sigma' \alpha' \beta} \}$. This behavior can be confirmed in Fig. \ref{err_vs_itr.eps}. As a result of the prediction, the numerical error in eigenvalue $e^{(n)}_{i}$ with respect to the Lanczos iterations becomes less than $10^{-13}$ at around 40 iterations, a third of that without the prediction when $n \gtrsim 72$ [Fig. \ref{err_vs_itr.eps}(a)]. In this region, as $\epsilon^{(n)}_{1}$ is comparable with $\epsilon^{\prime (n)}_{1}$, as shown in Fig. \ref{err_vs_itr.eps}(b), we find that the wavefunction prediction gives a nearly best-approximated eigenvector.

\subsection{Ground-state energy in the bulk limit under a fixed $m$}
Taking the double limit, namely, the number of iterations of calculations $n \rightarrow \infty$ and the number of maintained states $m \rightarrow \infty$ of the iDMRG, we can address the true physical quantity of the cylinder in the thermodynamic limit. In this and the next subsection, we show how to take the double limit correctly when estimating the ground-state energy per site of the cylinder. 

\begin{figure}[htb]
\centering
\includegraphics[width=8.5cm]{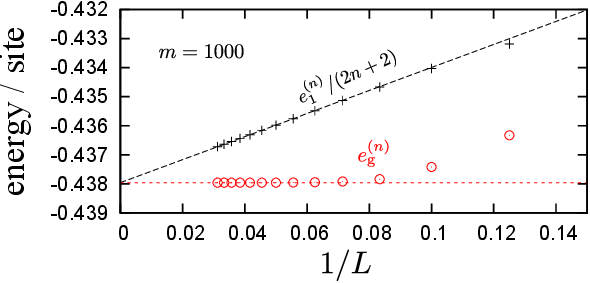} 
\caption{(Color online) Energy per site of finite systems $e^{(n)}_{1}/(2n+2)$ and the subtracted energies $e^{(n)}_{\rm g}$ of the YC8 kagome cylinder as a function of the reciprocal of cylinder length $1/L = (2n+2)/\ell$ with $m=1000$. }
\label{ene_vs_itr.eps}
\end{figure}

We first focus on the convergence of the energy per site with respect to $n$ under a fixed $m$. As shown in Step 6 of Sec. \ref{sec.p_idmrg}, we used the subtraction method~\cite{doi:10.1146/annurev-conmatphys-020911-125018} for suppressing edge effects to obtain the energy per site in the limit $n \rightarrow \infty$. If this treatment is suitable for accelerating convergence, $e^{(n)}_{\rm g}$ can rapidly converge to the energy per site of $\lim_{n \rightarrow \infty} e^{(n)}_{1}/(2n+2)$. As shown in Fig. \ref{ene_vs_itr.eps}, the energy per site had an almost linear dependence on $1/L$, depicted as the broken black line. We found that the energy per site $\lim_{n \rightarrow \infty} e^{(n)}_{1}/(2n+2)$ agreed with the convergent values of subtracted energies $e^{(n)}_{\rm g}=-0.43796022(2)$ up to $n=1200$, where the error was owing to the common cancellation of significant digits in the subtraction analysis. Using this subtraction method, we thus avoided a careful extrapolation of the energy per site with respect to the length of the cylinder. 

\subsection{Definition of truncation error in the parallel iDMRG}

\begin{figure}[htb]
\centering
\includegraphics[width=8.5cm]{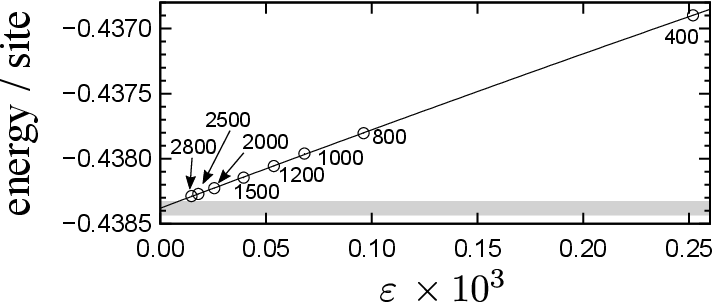} 
\caption{
Extrapolation of the ground-state energy per site with quadratic fits of $\varepsilon$ in the region $400 \leq m \leq 2800$. The numbers beside symbols show the values of $m$. The gray band indicates the estimated energy $-0.43838(5)$ in Ref.~\citen{PhysRevLett.109.067201}.}
\label{ene_vs_tru.eps}
\end{figure}

Following the above, to obtain the energy of the true ground state, we extrapolate $e^{(n)}_{\rm g}$ to the limit $m \rightarrow \infty$. In the parallel iDMRG, we define the truncation error $\varepsilon = \max[ \{ \varepsilon^{(n)}_{1} \} ]$ as
\begin{equation}
\varepsilon^{(n)}_{i} = 1 - \sum_{\beta=1}^{m} \left( \lambda^{(n)}_{i,\beta} \right)^2 
= (2-\epsilon^{\prime(n)}_{i})\epsilon^{\prime(n)}_{i}~ \label{eq.tru}
\end{equation}
and extrapolate $e^{(n)}_{\rm g}$ to the limit $\varepsilon \rightarrow 0$ because this truncation error is reduced by increasing $m$ and must be zero in the limit $m \rightarrow \infty$. The reasons for using $i=1$ and the maximum of $\{ \varepsilon^{(n)}_{i} \}$ are as follows: 
\begin{itemize}
\item[i)] The leftmost site of the cluster, represented by the renormalized Hamiltonian $H_i^{(n)}$, is fixed at the $i$th site irrespective of $n$ as shown in Fig.~\ref{fig:parallel_idmrg}, and the series of $\{ H_1^{(n)} \}$ have a chance of achieving the same boundary condition as the original Hamiltonian in Eq.~(\ref{eq.mpo2}).
\item[ii)] Reflecting the $\ell$-site period structure of the system, the value $\varepsilon^{(n)}_{1}$ has the periodicity with respect to $n$ shown in Fig.~\ref{err_vs_itr.eps}. We assume that the largest truncation error mainly determines the quality of the MPS. 
\end{itemize}
If the value of $\varepsilon$ is appropriate, we find that the expectation values fit well with the quadratic polynomial of $\varepsilon$ (which has a small quadratic dependence on $\varepsilon$) in the region $\varepsilon \ll 1$, as discussed in Ref.~\citen{PhysRevLett.99.127004}. As shown in Fig.~\ref{ene_vs_tru.eps}, the quadratic fit yields the extrapolated value $-0.43838(1)$, where the error is the standard deviation of the fit. The extrapolated value agrees with the reported values $-0.43836(2)$ and $-0.43838(5)$ up to $m=8000$~\cite{Yan03062011} and 16000\cite{PhysRevLett.109.067201}, respectively. 

\subsection{Bond strength of nearest-neighbor spins}
\begin{figure}[htb]
\centering
\includegraphics[width=8.5cm]{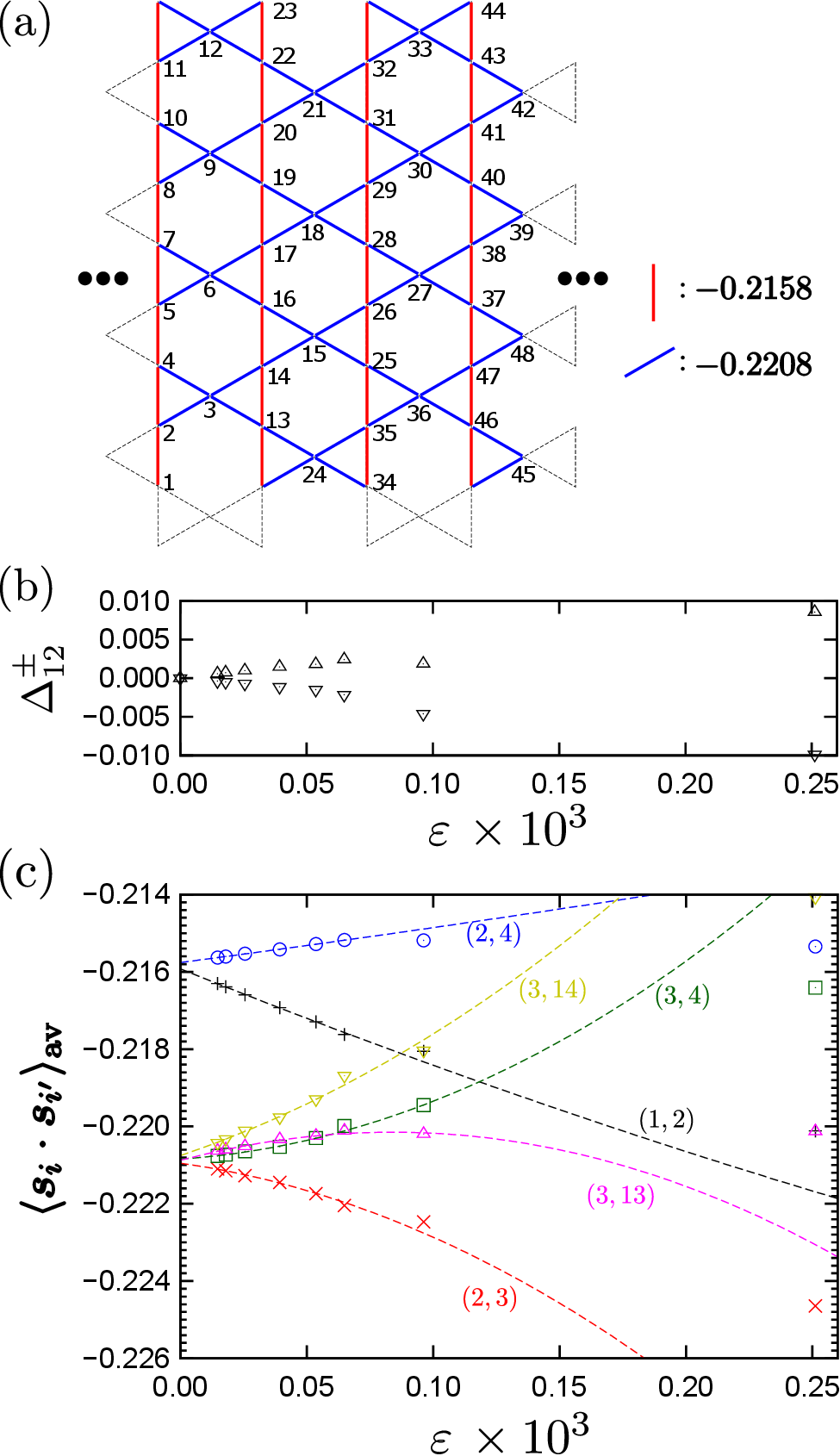} 
\caption{(Color online) (a) Configuration of bond strength in the ground state on the YC8 kagome cylinder. Numbers near lattice points indicate the order of sites for the MPS of the iDMRG. (b) Dependence of the difference between Eqs. (\ref{diff1}) and (\ref{diff2}) on the truncation error, where the triangles and inverted triangles represent $\Delta^{+}_{12}$ and $\Delta^{-}_{12}$, respectively. (c) Average values of bond strength $\langle {\bm s}_{i} \cdot {\bm s}_{i'} \rangle_{\rm av}$, where the broken lines are quadratic-fitted curves for data with $\varepsilon < 6 \times 10^{-5}$ in each pair of $(i,i')$. }
\label{bond_vs_tru.eps}
\end{figure}

To better understand the convergence behavior of the parallel iDMRG, we discuss the bond strength of the nearest-neighbor spins as a typical local observable. The parallel iDMRG can predict and assume a spatially uniform MPS with 12-site unit cells as in Eq. (\ref{eq.mps2}). Therefore, the correlation functions $\{ \langle {\bm s}_{i+12n} \cdot {\bm s}_{i'+12n} \rangle \}_{n} $ are equivalent to one another in our MPS, where the numbering of sites for the cylinder YC8 is shown in Fig. \ref{bond_vs_tru.eps}(a). Moreover, if the numerical calculations are executed exactly, the four correlation functions $\{ \langle {\bm s}_{i+3k} \cdot {\bm s}_{i'+3k} \rangle~|~0 \leq k \leq 3,~{\bm s}_{i+3k} = {\bm s}_{i+3k-12}~{\rm if}~\mod[i+3k-1,12] = \mod[i-1,12]+1 \}$ identically reflect the translational symmetry along the circumference of the cylinder. However, in our parallel iDMRG, these identities do not hold because of the finite-$m$ effect. Figure \ref{bond_vs_tru.eps}(b) shows the differences between bond strengths and the average value,
\begin{eqnarray}
\Delta^{+}_{ii'} & = & \max_k[ \langle {\bm s}_{i+3k} \cdot {\bm s}_{i'+3k} \rangle ] - \langle {\bm s}_i \cdot {\bm s}_{i'} \rangle_{\rm av}, \label{diff1} \\
\Delta^{-}_{ii'} & = & \min_k[ \langle {\bm s}_{i+3k} \cdot {\bm s}_{i'+3k} \rangle ] - \langle {\bm s}_i \cdot {\bm s}_{i'} \rangle_{\rm av}, 
\label{diff2}
\end{eqnarray}
where $\langle {\bm s}_i \cdot {\bm s}_{i'} \rangle_{\rm av}$ is the arithmetic average of $\{ \langle {\bm s}_{i+3k} \cdot {\bm s}_{i'+3k} \rangle \}_k$. The difference $\Delta^{\pm}_{12}$ approaches zero in the limit $\varepsilon \rightarrow 0$, and we can confirm the extrapolated values $| \lim_{\varepsilon \rightarrow 0} \Delta^{\pm}_{12} |$ by ensuring that the quadratic fits for data with $\varepsilon < 6\times10^{-5}$ are less than $1.0 \times 10^{-5}$. This behavior is consistent with the fact that the translational symmetry along the circumference must be recovered at the limit $m \rightarrow \infty$. Thus, we can focus on $\langle {\bm s}_i \cdot {\bm s}_{i'} \rangle_{\rm av}$ if we discuss the values at the limit $\varepsilon \rightarrow 0$. 

As there are two types of translational symmetry, we only estimate the set of bond strengths $\mathcal{B}=\{ \langle {\bm s}_{i} \cdot {\bm s}_{i'} \rangle_{\rm av}~|~(i,i')=(1,2),(2,3),(3,4),(2,4),(3,13)~{\rm and}~(3,14) \}$ to discuss the bond strength of nearest-neighbor spins on the cylinder YC8. Figure \ref{bond_vs_tru.eps}(c) shows $\langle {\bm s}_{i} \cdot {\bm s}_{i'} \rangle_{\rm av} \in \mathcal{B}$ versus $\varepsilon$. The values extrapolated to the limit $\varepsilon \rightarrow 0$ can be grouped into two values, $-0.2158(1)$ and $-0.2208(1)$. The configuration of the strength of nearest-neighbor spins corresponding to this result is shown in Fig. \ref{bond_vs_tru.eps}(a). 
A similar configuration result was reported for an XC8 kagome cylinder~\cite{Yan03062011}. 

\subsection{Spin-spin correlations and correlation length}
\begin{figure}[htb]
\centering
\includegraphics[width=8.5cm]{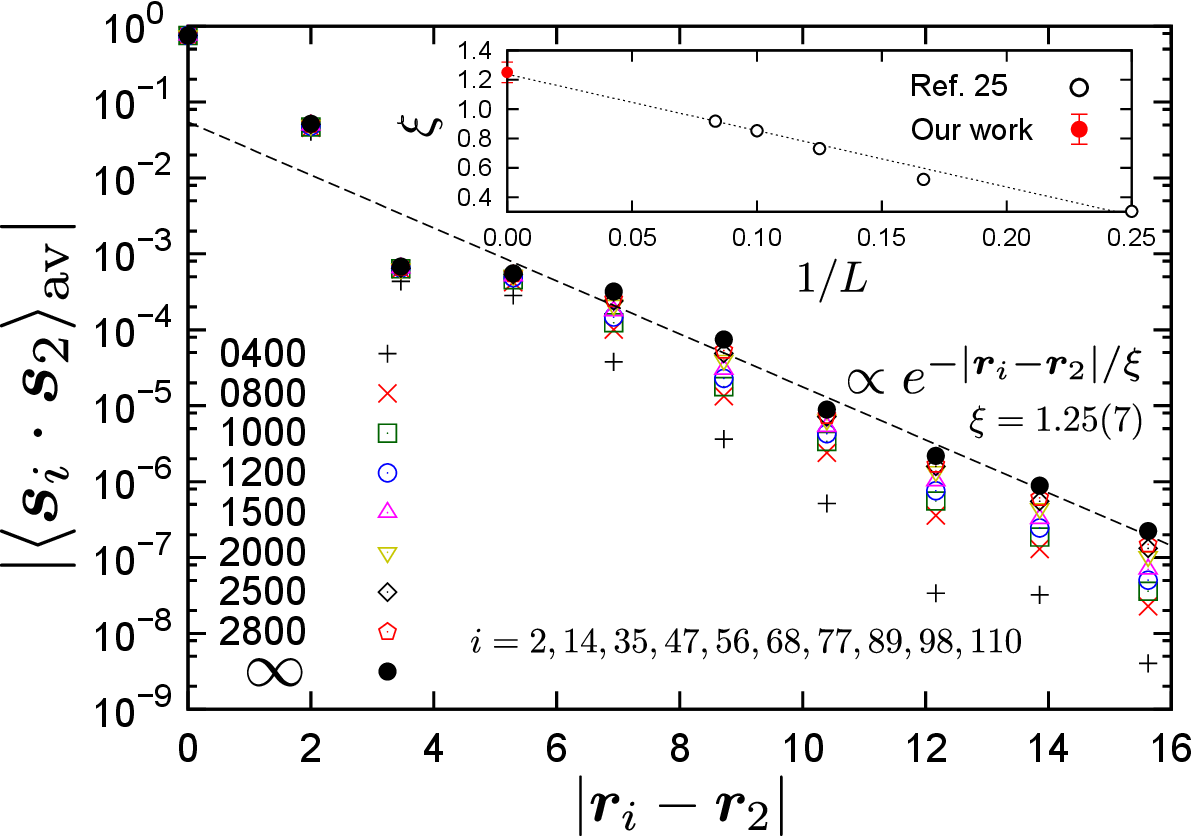} 
\caption{(Color online) Semilog plot of the absolute value of spin-spin correlation functions $\langle {\bm s}_{i} \cdot {\bm s}_{2} \rangle_{\rm av}$ along the cylinder YC8 with an exponential fit (dashed line). We used $m$ up to 2800. The numbers in the legend show the values of $m$. The closed black circles are the extrapolated values obtained using quadratic fits with respect to the truncation error, as discussed for Fig.~\ref{bond_vs_tru.eps}(c), and the fitting errors of $\langle {\bm s}_{i} \cdot {\bm s}_{i'} \rangle$ are smaller than the symbol sizes. The inset shows the comparison between the spin-spin correlation lengths estimated here and those reported in Ref.~\citen{PhysRevB.91.104418}, where the dotted line is the linear fit to data for $L=10$ and 12.}
\label{cf_vs_x.eps}
\end{figure}

As the final demonstration, we estimated the spin-spin correlation function $\langle {\bm s}_{i} \cdot {\bm s}_{i'} \rangle$ along the cylinder YC8 and determined its correlation length. We set $i' = 2$ and swept $i=2,14,35,47...$ along the axis of the cylinder [see Fig.~\ref{bond_vs_tru.eps}(a)]. The absolute values of the correlation function decayed exponentially with respect to distance $|{\bm r}_i-{\bm r}_2|$ between sites $i$ and $2$, as shown in Fig.~\ref{cf_vs_x.eps}. We fit the data for $|{\bm r}_i-{\bm r}_2| > 5$ with an exponential function $\propto e^{-|{\bm r}_i-{\bm r}_2|/\xi}$ and obtained the correlation length $\xi=1.25(7)$, where the error was the standard deviation of the fit. The length of the spin-spin correlation with finite length $L$ up to $12$ has already been evaluated by the non-abelian DMRG~\cite{PhysRevB.91.104418}. We confirmed that the correlation length $L \rightarrow \infty$ obtained by our parallel iDMRG agreed with the value extrapolated from the data for $L=10$ and $12$ with a linear fit. 

\section{Conclusions}
In this study, we investigated a parallel iDMRG method applied to 1D quantum systems with a large unit cell. This parallel iDMRG is based on Hida's iDMRG~\cite{Hida:JPSJ65} for 1D random quantum systems and a variant of McCulloch's wavefunction prediction~\cite{0804.2509}. The numerical efficiency of our parallel iDMRG was demonstrated for the spin-1/2 Heisenberg model on the YC8 kagome cylinder. Using the truncation errors proposed in this work, we succeeded in obtaining correct observables, including the ground-state energy per site, the bond strength on nearest-neighbor spins, and spin-spin correlation functions and their correlation lengths with the number of renormalized states $m$ up to $2800$, approximately a third (sixth) of the number of renormalized states in Ref.~\citen{Yan03062011} (Ref.~\citen{PhysRevLett.109.067201}). The wavefunction prediction increased the speed of the Lanczos methods in the our parallel iDMRG by approximately three times. This effectively reduced the numerical cost of the iDMRG. 

\begin{figure}[htb]
\centering
\includegraphics[width=8.5cm]{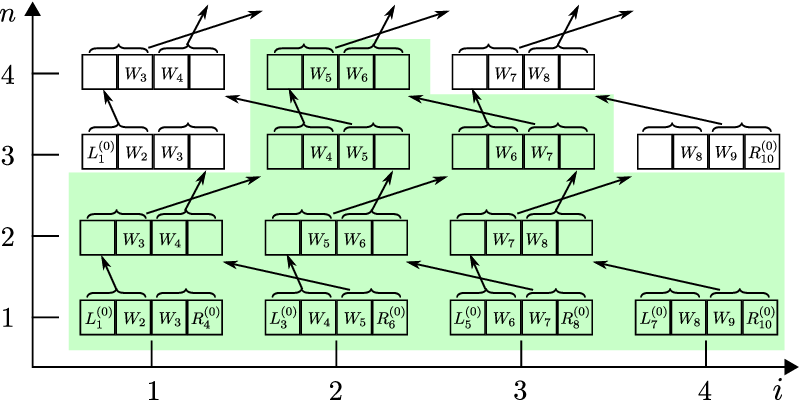} 
\caption{ (Color online) Schematic procedure of real-space parallel DMRG~\cite{PhysRevB.87.155137}, starting from Hida's iDMRG for random systems~\cite{Hida:JPSJ65}. }
\label{diagram3-8.eps}
\end{figure}

Several remarks are in order. First, Hida's iDMRG is intimately related to the real-space parallel DMRG~\cite{PhysRevB.87.155137}. Figure \ref{diagram3-8.eps} shows the entire picture of the real-space parallel DMRG, starting from Hida's iDMRG, where the diagrams, including overbraces and arrows, have the same meanings as those in Figs.~\ref{fig:Hida_idmrg} and \ref{fig:parallel_idmrg}. The region shaded in green is identical to Hida's iDMRG. In the procedures shown in Figure \ref{diagram3-8.eps}, the initial MPS is no longer needed to start parallel DMRG calculations.

Second, the physical background of wavefunction prediction in the iDMRG is understood well from the viewpoint of two-dimensional classical vertex models. By applying the quantum--classical correspondence discussed in Ref.~\citen{JPSJ.79.044001}, it can be easily shown that our parallel iDMRG algorithm is also applicable to analyses of 2D classical vertex models with arbitrary periodic structures along only the horizontal (vertical) direction. 

Third, our parallel iDMRG is compatible with other parallel algorithms, such as those used for parallelization over different terms in the Hamiltonian~\cite{chan2004} and the block diagonalization of a matrix with respect to the quantum number~\cite{Hager2004795,Kurashige2009}.  We expect that our parallel iDMRG and its extensions can be used in a variety of other quantum systems.

\section*{Acknowledgements}
The author thanks S. Onoda, T. Nishino, and S. Yunoki for fruitful discussions. This study was partially supported by JSPS KAKENHI Grant Numbers JP25800221 and JP17K14359. The computations were performed using the facilities at the HOKUSAI Great Wave system of RIKEN. 


\end{document}